%% file: zika.tex
\documentclass[conference]{IEEEtran}

\usepackage{booktabs}
\usepackage{balance}  
\usepackage{graphics} 
\usepackage{url}      
\usepackage{subfigure}
\usepackage{fixltx2e}
\usepackage{color}
\usepackage{float}
\usepackage{multirow}
\usepackage{hyphenat}
\usepackage[flushleft]{threeparttable}
\usepackage[normalem]{ulem}
\usepackage{array}
\usepackage{tabulary}
\usepackage{tabularx}
\usepackage{amsmath,array,graphicx}
\usepackage{kantlipsum}
\usepackage{float}
\usepackage{enumitem}

\usepackage{tikz}
\def\checkmark{\tikz\fill[scale=0.3](0,.35) -- (.25,0) -- (1,.7) -- (.25,.15) -- cycle;}
\newcolumntype{L}{>{\centering\arraybackslash}m{6cm}}
\newcommand\Tf{\rule{0pt}{2.8ex}} 
\newcommand\Ts{\rule{0pt}{2.2ex}}

\begin{document}

\title{Catching Zika Fever: Application of Crowdsourcing and Machine Learning for Tracking Health Misinformation on Twitter}

\author{\IEEEauthorblockN{Amira Ghenai}
\IEEEauthorblockA{University of Waterloo\\
Waterloo, Canada\\
Email: aghenai@uwaterloo.ca}
\and
\IEEEauthorblockN{Yelena Mejova}
\IEEEauthorblockA{Qatar Computing Research Institute\\
HBKU, Doha, Qatar\\
Email: ymejova@hbku.edu.qa}}

\maketitle

\begin{abstract}
In February 2016, World Health Organization declared the Zika outbreak a Public Health Emergency of International Concern. With developing evidence it can cause birth defects, and the Summer Olympics coming up in the worst affected country, Brazil, the virus caught fire on social media. In this work, use Zika as a case study in building a tool for tracking the misinformation around health concerns on Twitter. We collect more than 13 million tweets -- spanning the initial reports in February 2016 and the Summer Olympics -- regarding the Zika outbreak and track rumors outlined by the World Health Organization and Snopes fact checking website. The tool pipeline, which incorporates health professionals, crowdsourcing, and machine learning, allows us to capture health-related rumors around the world, as well as clarification campaigns by reputable health organizations. In the case of Zika, we discover an extremely bursty behavior of rumor-related topics, and show that, once the questionable topic is detected, it is possible to identify rumor-bearing tweets using automated techniques. Thus, we illustrate insights the proposed tools provide into potentially harmful information on social media, allowing public health researchers and practitioners to respond with a targeted and timely action.
\end{abstract}

\input{introduction}

\input{relatedwork}

\input{data}
\input{results}

\input{discussion}

\input{conclusion}

\section*{Acknowledgment}
We would like to thank Luis Fernandez-Luque, scientist, Qatar Computing Research Institute, William Schulz, research assistant, London School of Hygiene \& Tropical Medicine, Clarissa Simas, research assistant, London School of Hygiene \& Tropical Medicine and Per Egil Kummervold, Research Scientist, Norut Northern Research Institute for the fruitful discussions and health related expertise in the Zika rumor selection process.

\bibliographystyle{IEEEtran}
\bibliography{zika} 
\end{document}

%% file: introduction.tex

\section{Introduction}
\label{sec:introduction}

The information overload poses serious challenges to public health, especially with regards to infectious diseases. Similar to people's increased mobility, the availability and ubiquity of information facilitates the transmission of misinformation and rumors that can hamper the efforts to tackle a major public health crisis. With a continuous threat of digital ``wildfires'' of misinformation \cite{Webb2016}, health rumors are a worldwide and serious problem \cite{Fernandez-Luque2015}.

The complexity of dealing with communication during a health crisis is growing, as social media is playing a more prominent role. Social media, compared with traditional media, is harder to monitor, track and analyze. Public health institutions such as the World Health Organization (WHO) include social media as a crucial part in monitoring a health crisis \cite{WorldHealthOrganization2011}. However, guidelines and tools on best approaches to tackle this are not yet available. 

This paper proposes a suite of tools for tracking health-related misinformation, and describes a case study of tracking a health crisis, as discussed on Twitter. We provide a methodology for uncovering the streams of tweets spreading rumors about the 2016 Zika outbreak identified by the WHO. The Zika virus has been known for decades; it was discovered in Uganda in the 1940s and until recently it has been unnoticed. Things changed dramatically in 2015 when this mosquito-borne disease started to spread quickly across Brazil and then most of the American continent, becoming a major global health crisis. This crisis became more dramatic as the link between the Zika infection and serious brain malformation (i.e. microcephally) started to emerge. Furthermore, fears of a global pandemic started to emerge, since Zika is spread by a mosquito from the \textit{Aedes} family, which is present in many countries. Another source of concern were the Rio summer Olympics Games, which brought international travelers to the affected areas. As at the time there was no cure or vaccine for the Zika viral infection, communication with the public was one of the most important tools to control this outbreak. These communication efforts -- dealing with the detection and prevention of Zika, and also the reduction of mosquito breeding -- have been challenged by the appearance of rumors that, in the best of cases, were interfering with the public health campaigns \cite{rumorRisk}.

In particular, we track rumors outlined by the WHO (along with Snopes.com\footnote{http://www.snopes.com/}) in the stream of nearly 13 million tweets. We employ both automated LDA-based topic discovery as well as high-precision expert-led Information Retrieval approach to identifying the relevant tweets in this stream. Using crowdsourcing, we distinguish between rumor and clarification tweets, which we then use to build automatic classifiers. Here, we present in-depth temporal analysis of the found rumors, their origins, and interactions with informational sources. 

As discussed in the coming section, this work contributes to the literature a first application of the state-of-the-art social media analytic tools to the problem of health rumor tracking.

%% file: relatedwork.tex

\section{Related Work}
\label{sec:related}

This paper is the first of its kind to relate health informatics, machine learning and social network analysis to detect health-related rumors in a social media site. Below, we outline the most relevant recent work related to the detection and tracking of health-related attitudes, as well as misinformation in other non-health related fields. 

\textbf{Health-related attitudes on social media.} As social media captures daily thoughts and actions of millions of users, recent efforts have been made in extracting health-related behaviors and attitudes from the textual content of these websites. Recently, \cite{ginart2016drugs} build a classifier to detect the use of marijuana on Twitter. Attitudes toward legal drugs, including Xanax and Adderall, have also been studied by \cite{seaman2016prevalence}. Further, \cite{yang2013harnessing} use association mining to health communities to discover adverse effects of drug interactions. Health-related attitudes around food on Instagram have been mined by \cite{mejova2015foodporn}. These studies often combine the health expert knowledge with big data analytics (including machine learning, in the case of \cite{ginart2016drugs}), to provide insight into attitudes captured in social media interactions. Similar tools can be applied to non health-related rumor detection, as we discuss next.

\textbf{Rumor detection and analysis.} Separating newsworthy stories from misinformation and rumors across microblogging sites has been a popular research topic in recent years. 
Some focus on identifying information credibility of news propagated in social media  \cite{castillo2011information,leskovec2009meme,yang2012automatic,wu2015false}. 
Others focus on detecting misleading political memes, such as the ``Truthy'' service presented by Ratkiewicz et al.\ \cite{ratkiewicz2011truthy}, designed to detect fake political grass-roots movements (dubbed as ``astroturf''). 
Often, machine learning models based on features related to either users or the content of propagated messages are employed. For example, \cite{wu2015false,yang2012automatic} train a classifier to automatically detect the rumors on Sina Weibo microblog site\footnote{http://weibo.com}. They use Sina Weibo rumor-busting service\footnote{http://weibo.com/weibopiyas} as ground-truth data for the classification. Further, Jing Ma et al. \cite{ma2015detect}, use both Twitter and Sina Weibo as a case study in order to detect rumors. Using time series, they capture the temporal characteristics of the rumor detection features. Authors use already labeled datasets (Twitter \cite{castillo2011information}, Weibo \cite{wu2015false}) to train and evaluate the approach. Unlike these studies, we consider a more natural setting wherein no labeled data is available, and the documents associated with the rumors must first be found. In this respect, our approach is similar to Qazvinian et al.\ in \cite{qazvinian2011rumor} who use content-based, network-based, and Twitter-specific features in order to track urban legends. They build Bayes classifiers using engineered features and then learn a linear function of these classifiers for rumor retrieval and classification. 


\textbf{Health Related misinformation detection}. WHO's white paper on Risk communication in the context of Zika virus urges to ``build capacity to quickly transform new information
into usable, culturally-appropriate and easily understood risk communication resources that can be disseminated on multiple platforms'', including social media \cite{world2016risk}. However, little work thus far concentrated on detecting and tracking health-related rumors. 
Recently, Kostkova et al.\  \cite{kostkova2016vac} created the ``VAC Medi+ board'' online interactive visualization framework integrating heterogeneous real-time data streams with Twitter data. They track the spread of vaccine related information on Twitter and the sources of information spread. 
A potential framework to engaging expert knowledge in a real-time crisis, including health-related, situation is described in Imran et al.\ \cite{imran2014aidr}, where content is selected to be annotated via crowdsourcing into pre-defined classes. These can then be used to train a classifier, and update it as necessary with active learning data selection. 
Work most relevant to the current study is by Dredze et al.\ \cite{dredze2016zika}, who analyzed the characteristics of nonscientific claims about vaccine misconceptions by the vaccine refusal community. Specifically, the authors analyzed the two most prominent misleading theories about Zika vaccination in Twitter using ``supervised machine learning technique'' (although these were not explained in the manuscript) and observed the effect of vaccine-skeptic communities over other users' vaccination opinion.
While \cite{dredze2016zika} look at two Zika vaccine related memes, in this work we propose a more general methodological pipeline to track health-related rumors. Taking Zika as a case study, with the help of health professionals we expand the list of rumors to six, and examine the behavior of rumor as well as clarification efforts. Below we describe in detail our contributions to the text classification task via novel health-related features, and the application of crowdsourcing to fine-grained document labeling.


%% file: data.tex

\section{Data Collection}
\label{sec:data}

\begin{figure}[t]
    \centering
    \includegraphics[width=\linewidth]{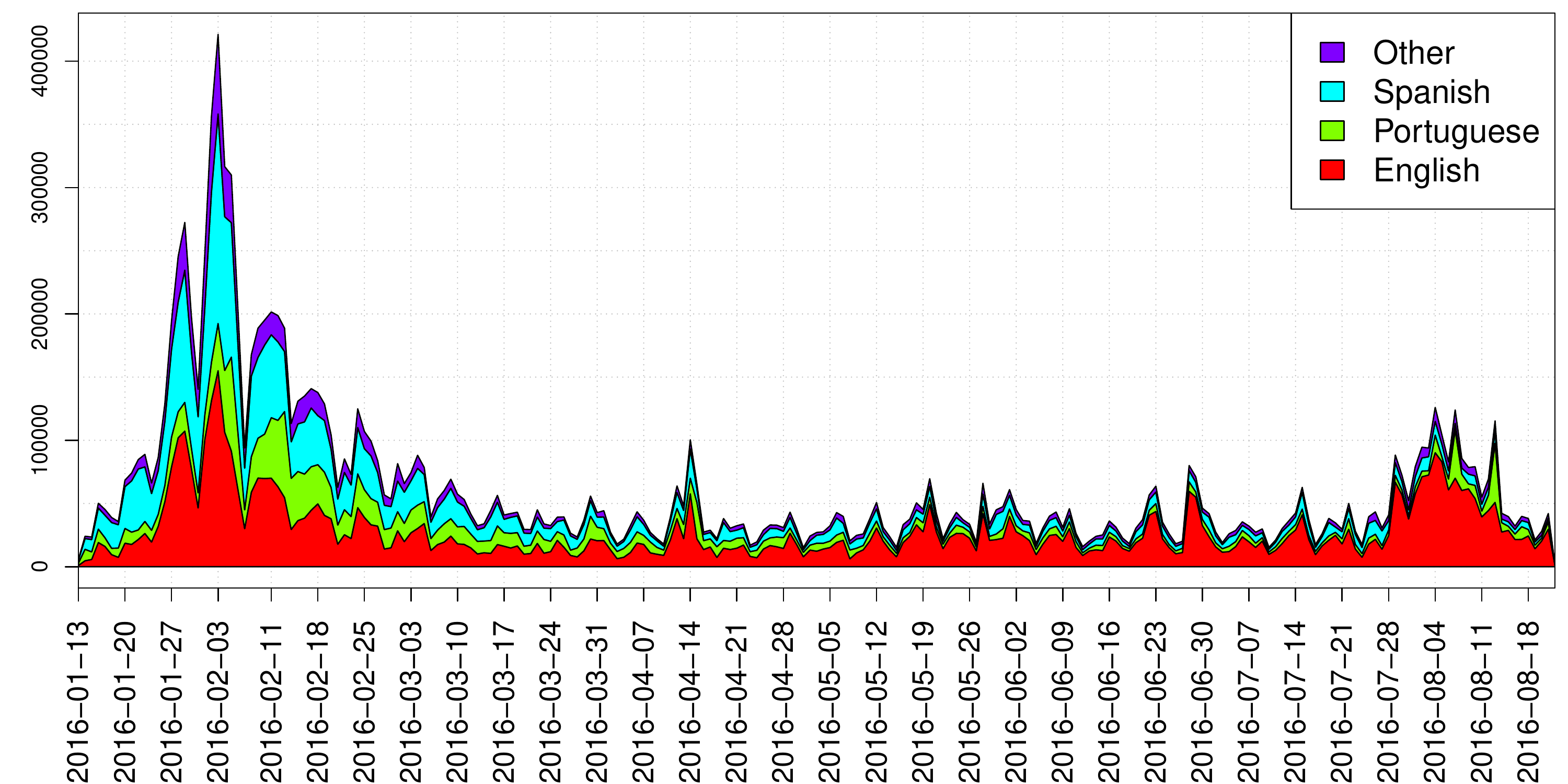}
    \caption{Zika-related Twitter data volume, separated by language.
    }
    \label{datavolume}
\end{figure}

The data was collected using the Artificial Intelligence for Disaster Response (AIDR)\footnote{http://aidr.qcri.org/} platform, which taps into Twitter Streaming Application Program Interface (API). The keywords' list contained the following (searched as quoted strings): zika, microcefalia, microcephaly, \#zika, zika virus, \textit{Aedes}, zika fever, \textit{Spondweni} virus, \textit{Aedes albopictus}, 
\textit{maculopapular} rash. We aimed to cover both everyday wording as well as medical jargon which may be associated with the topic. Furthermore, "zika" word is used by all English, Portuguese, and Spanish, which are the major languages of populations affected.
The resulting collection of 13,728,215 tweets spans January 13 - August 22, 2016, and includes the peak of interest in Zika (in early February) and the Olympic Games in Brazil (August 5-21). Figure \ref{datavolume} shows the volume of the data by language. 

Since no language restriction was imposed during data collection (besides some bias English keywords introduced), we captured a plurality of languages, with three dominant ones which represent more than half of the dataset -- English, Spanish and Portuguese (46\%, 27\% and 17\% respectively). The language was determined using the language tag in the meta-data of the tweet provided by the Twitter API. Table \ref{table:language_stats} summarizes the global statistics of these languages distributions. It illustrates the international nature of the Zika crisis, with each language identifying a population and its diasporas affected. In this work we focus on the English data, and discuss the future work involving other languages below.

\begin{table}[t]
\centering
\caption{Data statistics by main language groups.}
\label{table:language_stats}
\begin{threeparttable}
\begin{tabular}{|l|r|r|c|l}
\cline{1-4}
\textbf{Language}\Tf & \textbf{Total tweets} & \textbf{Users} &  \textbf{Tweets/user} &  \\ \cline{1-4}
English         \Tf  & 6,267,173               & 1,318,293    & 4.75 &  \\ 
Spanish         \Tf  & 3,689,292               & 727,105    & 5.07 &  \\ 
Portuguese       \Tf & 2,296,611               & 623,968    & 3.68 &  \\ 
Other           \Tf  & 1,475,139               & 593,221    & 2.49 &  \\ \cline{1-4}
Total            \Tf & 13,728,215              &  3,262,587 \tnote{$\star$}  & 4.21 &  \\ \cline{1-4}
\end{tabular}
\begin{tablenotes}
\item  ${}^\star$ or 2,546,851 unique users
\end{tablenotes}
\end{threeparttable}
\end{table}

To understand the geographic distribution of tweets, we use several sources to geo-locate them. We begin with the GPS attributes (\textit{latitude}, \textit{longitude}) of the tweet meta-data and convert them to the corresponding country name using World Borders API\footnote{http://thematicmapping.org/downloads/world\_borders.php}. As 99\% of tweets have missing GPS attributes, we look at location names in the \emph{place} attribute of the tweet and convert them to exact country names. In case there is no mentioned location, we assume that the tweet location is where the user is located and get the corresponding location from the user's profile. It is worth mentioning that users' locations are messy, as they are written by the users. Thus, we use Yahoo Placemaker API\footnote{http://www.programmableweb.com/api/yahoo-placemaker} to map the users' place fields to GPS locations. Finally, in case no user location is mentioned, we get the tweet location by looking into other already identified location tweets tweeted by the same user (resorting to this noisy approach only if all other geo-location attempts fail). Implementing these steps, we achieve 68\% coverage for English data. The top locations in decreasing number of tweets are the United States, United Kingdom, India, Canada, Nigeria, and Brazil, indicating a highly international data.




\section{Rumor Selection}

Our rumor selection process begins with a reliable list of information from trusted sources. We chose the WHO website as an authority for detecting and verifying rumors about Zika. As Zika was spreading further in the world, WHO provided a source listing major international rumors and misinformation about the virus. At the time of writing, WHO website \cite{WHO2} listed 8 statements debunking ongoing rumors. Out of these, 4 were unsuitable, as they were not topically cohesive. For instance, one explained ``Fish can help stop Zika'', but did not explicitly state what the rumor which this statement would debunk would be. The process in understanding whether a topic is a rumor involved writing out an unambiguous description of the rumor, such that a tweet may be easily classified as being one. If no such description could be written, the topic was discarded. Additionally, we employed Snopes.com
, which is an online authority for detecting and verifying rumors in social media, emails and other online networks \cite{nourbakhsh2015newsworthy}, based on expert sourcing. From its Zika-related articles, we selected rumors which are not specific to the US. The final list of rumors, shown in Table \ref{Zika-rumors-description}, along with example tweets which propagate it, includes a total of 6 Zika rumor stories (4 from WHO and 2 from Snopes). 

Note that the selection of these Zika rumor topics was supervised by health experts (acknowledged below) in order to insure the coverage of the most important and influential topics related to the Zika outbreak. 

\begin{table}[t]
\caption{Zika rumor descriptions and example tweets. First four come from WHO and last two from Snopes.}
\label{Zika-rumors-description}
\begin{tabular}{|p{3cm}|p{4.5cm}|}
\hline 
 \multicolumn{1}{|c|}{\textbf{Rumor Description \Tf }} & \multicolumn{1}{c|}{\textbf{Example tweets}} \\ \hline \hline
 R1) \Tf Zika virus is linked to genetically modified mosquitoes & \textit{BIOWEAPON! \#Zika Virus Is Being Spread by \#GMO \#Mosquitoes Funded by Gates!} \\
 \hline
R2) \Tf Zika virus symptoms are similar to seasonal flu & \textit{The affects of Zika are same symptoms as the Common Cold. \#StopSpreadingGMOMosquitos} \\ \hline
R3) \Tf Vaccines cause microcephaly in babies & {\textit{Government document confirms tdap vaccine causes microcephaly.. https://t.co/4ZVLbaabbG}} \\ \hline
R4) \Tf Pyriproxyfen insecticide causes microcephaly & \textit{"Argentine and Brazilian doctors suspect mosquito insecticide as cause of microcephaly"} \\ \hline
R5) \Tf Americans are immune to Zika virus  & \textit{Yup and Americans R immune to Zika, so why fund a response to it?} \\ \hline
R6) \Tf Coffee as mosquito-repellent to protect against Zika & \textit{Bring on the Cuban coffee. Say Goodbye to Zika mosquitoes. Dee Lundy-Charles Fredric Sweeney Joshua Oates Laure... http://fb.me/tArL595b} \\ \hline
\end{tabular}
\end{table}

%% file: results.tex
\section{Rumor Tracking}
\label{sec:results}

We first attempt to discover these rumors using an automated technique, as topic discovery has been used to identify rumors \cite{ma2015detect,wu2015false} in social media. We train a Latent Dirichlet Allocation (LDA) \cite{blei2003latent} model on the english-language tweets, which then produces $n$ ``topics'', grouping words which appear in similar contexts together in a topic. 
However, after a manual examination of $n$ (varying from 5 to 50) topics, we do not find any topics pertaining to the above selected rumors. The vast majority of topics were informational, following by spam and jokes. Thus, we illustrate the necessity of incorporating expert knowledge in order to achieve a high-precision view of the data for our purpose.


\subsection{Query Construction}

We consider the task of extracting tweets relevant to our rumors as a standard Information Retrieval task. We first index the collected tweets using Indri\footnote{http://www.lemurproject.org/indri.php}, and submit a set of handcrafted interactively designed search queries (similarly to \cite{qazvinian2011rumor}). Each query is a boolean string consisting of a list of keywords that best describe the rumor. These keywords are first identified, then connected using the AND, OR and NOT operators.  Each keyword is then replaced with a series of possible synonyms and replacements, all connected via the OR operator. For instance, consider the rumor saying that ``vaccines cause microcephaly'' (R3). Transforming this story to a query language would include several common ways of referring to vaccines, as shown in Table \ref{query-table}. The queries are hand-crafted over at least 3 iterations of labeling the top 10 returned results.

Designing the queries to extract the tweets was not a trivial task. One of the challenges is that many medical term synonyms needed to be added to the query to get the highest coverage. We did not rely on automatic query expansion techniques such as Pseudo-relevance feedback as these automatic algorithms perform well in medical articles and not in informal unstructured text such as Twitter messages \cite{queryExpansion}. Additionally, we added words that distinguish general information tweets from rumors. For example, in R2, to distinguish a rumor from a general information, we need to add (NOT rash) to the query because this is the symptom that differs between Zika symptoms and the seasonal flu ones. 

\begin{table}[t]
\caption{Rumor queries and the number of tweets retrieved.}
\label{query-table}
\begin{tabular}{|l|p{5.2cm}|r|}
\hline
\multicolumn{1}{|c|}{\textbf{No}} & \multicolumn{1}{c|}{\textbf{\Ts Regular Expression Query}} & \multicolumn{1}{c|}{\textbf{\# tweets}} \\ \hline \hline
R1 \Tf & genetically $\mid$ GMO & 73,832 \\ 
R2 & (symptom \& (flu $\mid$ cold)) \& (not(rash)) & 469 \\ 
R3 & (tdap $\mid$ MMR $\mid$ Measles $\mid$ Mumps $\mid$ Rubella) \& vaccine \& microcephaly) $\mid$ (vaccine \&(cause $\mid$ link $\mid$ relate) \& microcephaly) & 4,329 \\ 
R4 & (montsanto $\mid$ pesticide $\mid$ pyriproxyfen $\mid$ insecticide) \& microcephaly & 10,389 \\ 
R5 & american \& immune & 351 \\ 
R6 & ((coffee $\mid$ java $\mid$ jive) \& (repellent $\mid$ protect)) \& (java \& jive) \& (coffee \& mosquito)) & 202 \\ \hline
Total & - \Ts & 89,572 \\ \hline
\end{tabular}
\end{table}

The final retrieval resulted in 89,572 tweets varying greatly by rumor, with a maximum of 73,832 to 202 (Table \ref{query-table}). These tweets, however, still may contain false positives, tweets that match the query but are not a rumor. For example, the following tweets are all about vaccines and microcephaly in babies (R3). The first tweet is stating that Zika vaccine causes microcephaly \textbf{(rumor)}, but the second tweet clarifies that there is no evidence suggesting Zika vaccine causes microcephaly \textbf{(clarification)}, and the third does not mention anything specific about the relationship between Zika vaccine and microcephaly \textbf{(other)}.  \\ 

\indent
\textbf{(rumor)} \textit{Government document confirms tdap vaccine causes microcephaly.. https://t.co/4ZVLbaabbG}

\indent
\textbf{(clarification)} \textit{Anti-vaccination extremists falsely claim that Tdap \#vaccine causes microcephaly suspected to be caused by.. https://t.co/yvfHlAFKhw} 

\indent
\textbf{(other)} \textit{No cure, no vaccine for a virus that scientists believe to cause microcephaly! \#microcephaly \#ZikaVirus  https://t.co/EuG9b1AJVw}\\

In the coming section, we explain the approach we take in order to distinguish between the three different types of information available in our data-set.

\subsection{Crowdsourced Annotation}

To annotate the tweets as to whether they are indeed rumors, we employ the crowdsourcing platform ``Crowdflower''\footnote{http://www.crowdflower.com/}. Previous studies have shown that using crowds (anonymous workers) for health-related annotation is an effective way to label large amounts of data without employing experts \cite{yu2013crowdsourcing, zhai2013web}. We begin by creating a task for each topic with clear instructions on the labeling of the tweets as either supporting the rumor (by outright statement or ambiguity), debunking the rumor (by clarification), or doing neither. Also for each task we create a set of no fewer than 20 ``gold standard'' tweets (those with known classifications) in order to test the quality of annotations throughout the jobs. If an annotator did not pass the threshold of 70\% accuracy, he/she would be banned from the task and the annotations would be discarded. Each tweet was labeled at least 3 times and a majority vote determined its classification.

The tweets were first de-duplicated by stripping tweet-specific elements such as RT (standing for ``re-tweet"), special characters, and mentions, such that only one copy of each tweet was to be labeled. A maximum of 1,000 tweets were annotated per rumor. For those which had more than 1,000 unique tweets (R1 and R4), we first selected 700 most re-tweeted tweets, and sampled 300 from the rest. After the labeling of these unique ones, the label was then propagated to the duplicates within the set. 

\begin{table}[t]
\caption{Crowdflower label statistics of unique tweets in each category (propagated labels to duplicates in parentheses).}
\label{table:labels}
\begin{center}
{\fontsize{8pt}{12pt}\selectfont
\begin{tabular}{|l|rrrr|}
\hline
 & \textbf{Labeled} & \textbf{Rumor} & \textbf{Clarification} & \Ts \textbf{Other} \\ \hline
R1 & 1,000 (42,432) & 253 (11,773) & \Tf 50 (1,912) & 697 (28,747) \\
R2 & 302 (469) & 217 (348) & 71 (100) & 14 (21) \\
R3 & 796 (4,329) & 478 (2,853) & 88 (846) & 230 (630) \\
R4 & 1,000 (8,085) & 749 (5,586) & 221 (2,338) & 30 (161) \\
R5 & 131 (351) & 17 (22) & 99 (17) & 15 (312) \\
R6 & 114 (202) & 72 (129) & 5 (25) & 37 (48) \\\hline
\end{tabular}}
\end{center}
\end{table}

Table \ref{table:labels} shows the distribution of classes for the six rumors, with the number of tweets with propagated labels in parentheses. Although the queries were hand-crafted to capture rumors, only 51\% of final tweets were rumors (an average percentage across topics, such that no one topic dominates the statistic), and 15\% clarifications, attempting to debunk these rumors. The annotator agreement (as measured in label overlap) ranged between 76\% (R2) and 93\% (R5) with an average of 87.7\%, indicating the task differs in difficulty, but is overall clear to the annotators. 

\subsection{Temporal Tracking}
Next, we examine the ``paths'' these rumors have taken in the story line of Zika in our dataset. Figure \ref{rumorvolume} illustrates the bursty nature of these rumors. The plots also show Pearson product-moment correlation $r$ between the rumor and clarification volumes. For R4,5,6, the volume of clarification corresponds rather closely to that of the rumor with $r$ of around 0.5. However, R1,2,3 display a mismatch between clarification attempts and the rumors. We define the ``origin'' tweets for rumors or clarifications as the most prominent tweets at that time for the corresponding class and we explain Figure \ref{rumorvolume} in details as follows: 

\begin{figure}[h]
    \centering
    \includegraphics[width=\linewidth]{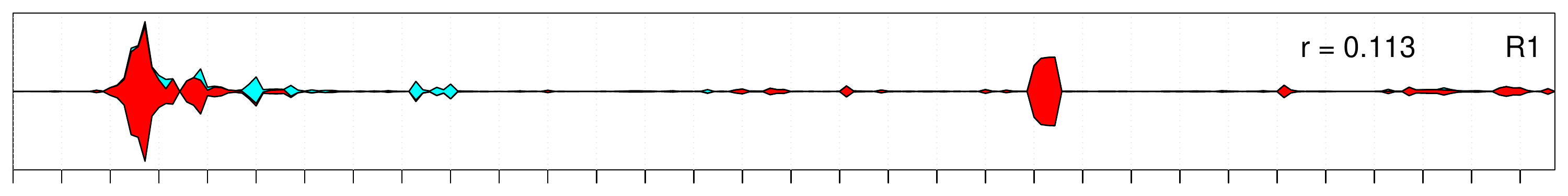}\\
    \vspace{0cm}
    \includegraphics[width=\linewidth]{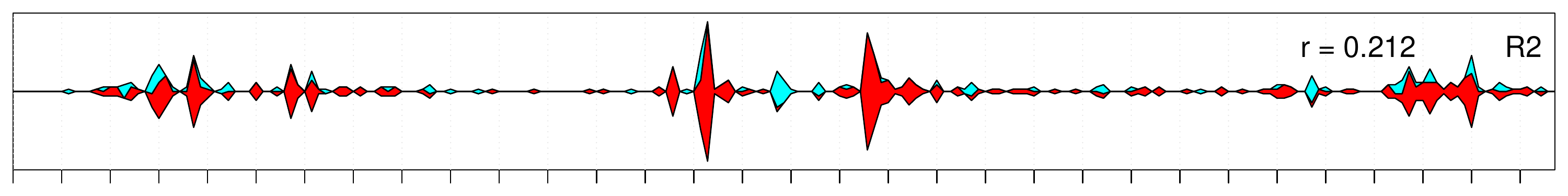}\\
    \vspace{0cm}
    \includegraphics[width=\linewidth]{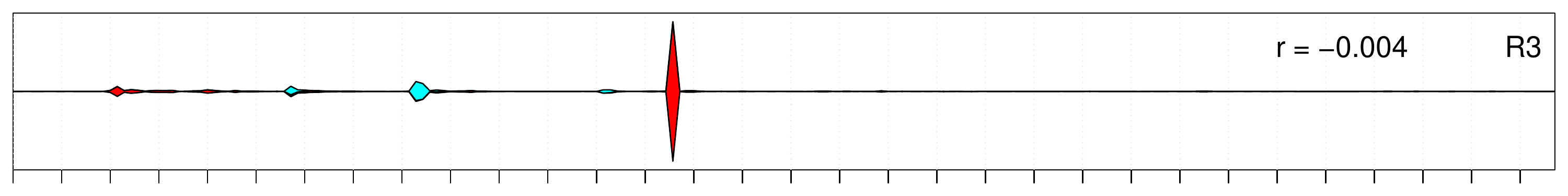}\\
    \vspace{0cm}
    \includegraphics[width=\linewidth]{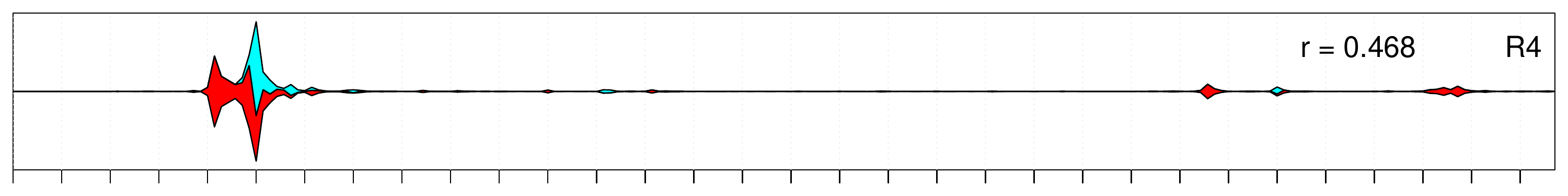}\\
    \vspace{0cm}
    \includegraphics[width=\linewidth]{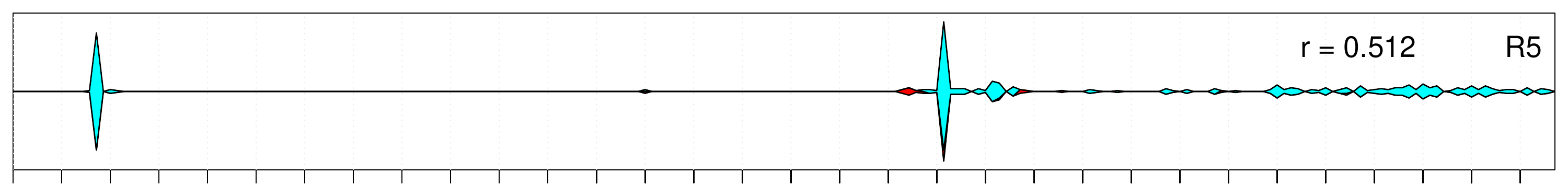}\\
    \vspace{0cm}
    \includegraphics[width=\linewidth]{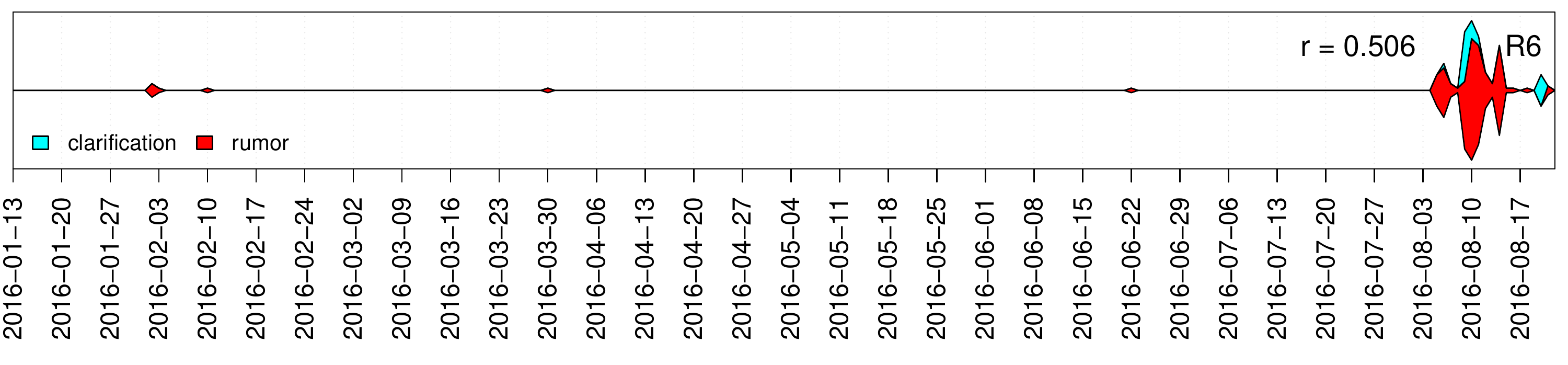}
    \caption{Volume of the six rumors and their clarifications, along with the Pearson product-moment correlation $r$ between the rumor and clarification volumes.}
    \label{rumorvolume}
\end{figure}
 
\textbf{The case of mutant mosquitos:} the case of R1 is interesting, since it carries over a concern about the dangers of genetically modified organisms (GMO) which has been popular for several years. For instance, the spike in July was due to an article published on The Real Strategy website claiming a link between ``chemical exposure'' and microcephaly\footnote{For more on this rumor see http://www.huffingtonpost.com/entry/zika-monsanto-pyriproxyfen-microcephaly\_us\_56c2712de4b0b40245c79f7c}, which gained thousands of retweets within days. However, without any interference that we detected from authoritative sources, the rumor quickly died out.

\textbf{Have you got Zika?} Flu and cold are very common diseases, therefore confusion between flu or cold and Zika might pose a serious problem for health authorities. This case is addressed in R2. Often the tweets appear to be jokes of users who feel flu-like symptoms such as:

\vspace{0.2cm}\noindent\hspace{0.1\linewidth}\begin{minipage}{0.8\linewidth}
\textit{RT @arzel: my friend had a small cold and I caught him googling ``zika virus symptoms''}
\end{minipage}\vspace{0.2cm}

\noindent Thus, although there are regular tweets on the true symptoms of Zika, there is a large proportion among these tweets that are jokes or lighthearted statements. 

\textbf{The killer vaccines:} Similar to R1, R3's peak originated in April with an article on another advocacy website www.march-against-monsanto.com (which argues that Monsanto, an agricultural biotechnology corporation, threatens the environment and the farmers) titled ``1991 Government Document Confirms Tdap Vaccine Causes Microcephaly''\footnote{http://www.march-against-monsanto.com/1991-government-document-confirms-tdap-vaccine-causes-microcephaly/}. The article was readily believable to people who already view Monsanto negatively and might be spread by pharmaceutical companies to create an opportunity to sell new Zika virus vaccines as Dredze et al. \cite{dredze2016zika} suggested in his paper. The post happened after a major WHO campaign in February and March saying ``No evidence that vaccines cause microcephaly''\footnote{https://twitter.com/WHO/status/708317001366806528}. Interestingly, the April spike receded just as quickly without any clarifications from authoritative sources.

\textbf{Pesticides, immunities and coffee grounds:} Others, however, did have a strong interaction between the rumor and a quick reaction with clarifications. For instance, the most retweeted stories of R4 are those coming from mainstream media including CNN and WHO stating there is ``No link between pesticide and microcephaly''. At the top three of R5 are stories on the ``crazy and dangerous story [that] Americans are immune to Zika'' and links to the debunking website Snopes. Similarly to R2, in R6 is a case of hyperbole and exaggeration of a story saying mosquito larvae do not thrive in coffee-infused water, which was turned into sensationalist tweets claiming ``Could Coffee Be the Answer in the Fight Against Zika Mosquitoes?'', but which still linked to the original correct information.

Thus, we show the varied nature of the rumors in the Zika stream. Those which were accompanied with mainstream coverage quickly decreased (R4-6), but even those which originated from the websites of various advocacy groups and were not met with official response were also short-lived (R1, R3). The longer-lived one is the one which concerned the daily occurrences (having a flu R2 or, possibly, coffee R6) which propagates in the Twitter lore.

\subsection{Rumor Classification}

\begin{table*}[t]
\centering
\caption{Automatically extracted features of tweets potentially belonging to a rumor.}
\label{features-table}
{\fontsize{8pt}{9pt}\selectfont
\begin{tabular}{|c|l|l|}
\hline
\textbf{ Scope } \Tf      & \textbf{Feature}   & \textbf{Description}                                              \\ \hline \hline
{Twitter} & IS RETWEET \Tf     & Is a retweet; contains RT    \\
& FOLLOWING          & The number of people the user is following                        \\ 
   & FOLLOWERS          & The number of people following the user                           \\ 
                          & STATUS\_COUNT      & The number of tweets at posting time                              \\ 
                          & AGE                & The time passed since the author registered his/her account, in days \\
                           & HAS MENTIONS       & Mentions a user, eg: @CNN                                         \\ 
                          & HAS HASH TAG      & Contains hash\_tags                                               \\ 
                          &COUNT HASH TAG & Count total number of hash\_tags
\\                          & DAY WEEKDAY        & The day of the week in which the tweet was written                \\ 
& COUNT URLS         & Count total number of URLs in text                                      \\  
 & COUNT RT          & Count total number of Retweets                            \\ & COUNTRY & The country the tweet was originated from  \\
 {Sentiment}  & SENTIMENT SCORE    & sentiment score value \cite{lowe2011scaling} \\
                          & POSITIVE/NEGATIVE WORDS     & The number of positive/negative words in text                          \\     
                          &EMOTICONS POS/NEG& Count total number of positive and negative emoticons in text\\
{Linguistic}                                            
                          & QUESTION MARK      & Contains question mark '?'                                        \\
                          & EXCLAMATION MARK   & Contains exclamation mark '!'                                     \\
                         
                          & WORDS COUNT        & Count total number of words in text                               \\ 
                         & COUNT SENTENCES & Count number of sentences\\
                          & CHAR COUNT         & Count total number of characters in text                          \\ 
                          & UPPER COUNT        & Count total number of upper case letters                                       \\ 
                          & PERCENTAGE UPPER & The percentage of upper case characters\\
                          & PERCENTAGE UPPER/LOWER & The percentage of upper and lower case characters \\ 
                         
                          & MULTIPLE QUES/EXCL & Contains multiple questions or exclamation marks                  \\ 
                        &COUNT NOUN   &Count total number of nouns in text\\
						&COUNT ADVERB       &Count total number of adverbs in text\\
						&COUNT ADJECTIVE   &Count total number of adjectives in text\\
						& COUNT VERB        &Count total number of verbs in text\\
						&COUNT PRONOUN      &Count total number of pronouns in text\\
                          & HAS PRONOUN 1    & Contains a personal pronoun in 1th person                         \\ 
                          & HAS PRONOUN 2    & Contains a personal pronoun in 2nd person                         \\ 
                          & HAS PRONOUN 3    & Contains a personal pronoun in 3rd person                         \\     
 {Readability} & COMPLEX WORDS       & Count total number of complex words in text                        \\
  &READABILITY SCORES  & Flesch, Automated, Flesch\_Kincaid, Gunning, and SMOG \cite{readabilityscore}\\
   &COUNT NOT WORD2VEC  & Count total number of words not in ``word2vec'' Google News vocabulary \\
   & AVG SYLLABLES & The Average number of syllables per word in text\\
   {Medical}& MEDICAL\_LEXICON & Count number of words in the medical lexicon \\ 
   & WIKIPEDIA DOMAIN & Count number of URL domains mentioned in the wikipedia web pages\\ & ADVOCACY       & Count number of URLs belonging to advocacy domains      \\  & NEWS & Count number of URLs belonging to news domains\\
    & SOCIAL & Count number of URLs belonging to social media domains\\
     & INFORMATIVE & Count number of URLs belonging to informative/trusted domains\\
                             \hline
\end{tabular}}
\end{table*}

Next, we turn to the supervised methods which have been proposed in previous work on news in social media that seek to establish the level of credibility of information automatically by observing specific features extracted from the social media. For instance, Castillo et al.\ \cite{castillo2011information} and Qazvinian et al.\ \cite{qazvinian2011rumor} suggested that the best features to assess the credibility of news topics are those that look into the user, message and topic features. Inspired by these works, we build a set of features in order to assess their power in automatically distinguishing rumors from non-rumors.

Gathering all the relevant tweets to the topics in Table \ref{rumorvolume}, results in a total of 56,985 tweets. Later, we filter tweets that are exact duplicates (tweets sharing exact similar information including text, urls, hashtags, and mentions) as the presence of the duplicates might influence the precision and recall values, resulting in a total of 26,728 tweets with human-assigned labels. We group the labels used in Table \ref{rumorvolume} such that we consider a rumor as the tweet that has been labeled by Crowdflower users as ``rumor'' (32\% - 8,488 tweets) and a non-rumor as the tweet that has either been labeled as ``clarification'' or ``other'' (68\% - 18,240 tweets). Note that we cannot consider ``clarification'' class alone, as it is vastly under-represented in our data (in part due to our focus on retrieving rumors in the previous steps).

The feature set is listed in Table \ref{features-table} and consists of 48 features grouped into five categories. The first three categories (Twitter, sentiment and linguistic features) have been previously implemented in news credibility detection, whereas the last two (readability and medical features) are new to this work:

\begin{description}[align=left]
\item [Twitter features] As \cite{castillo2011information} use Twitter features to define credibility in news topics, we build 18 similar features including the number of retweets, number of users followers and following, the presence of hashtags and mentions, the user's number of tweets, etc.

\item[Sentiment features] We consider five measures of emotional state in our dataset: count of positive/negative words, count of positive/negative smileys and sentiment score representing the strength of sentiment \cite{lowe2011scaling}.

\item [Linguistic features] We also introduce measures to characterize different linguistic styles in Twitter text \cite{castillo2011information}. We compute 17 different linguistic styles e.g: count adjectives, adverbs, pronouns, sentences, upper and lower case characters. 


\item[Readability features] Authors in \cite{readabilityscore} defined the readability score as a measure of how easy it is to understand a piece of text. We introduce a set of tweet text readability measures with the intuition that more readable information are more credible. We implemented the pre\-defined readability scores by \cite{readabilityscore} (Flech, automated, Flesch\_kincaid, Gunning, and SMOG scores) in addition to computing the number of complex words and average number of syllables per word. Moreover, we counted the number of words not in word2vec news vocabulary which may signal slang language \cite{word2vec}.

\item [\textbf{Medical/Domain features}]
We define specialized features in the medical domain by focusing on the medical lexicon of tweets and the reliability of sources shared using URLs. First, we build a medical lexicon\footnote{Available at http://bit.ly/2m56t0w} which signals how many medical terms are used in the tweet. Prior studies \cite{friedlin2010evaluation} showed that Wikipedia is a reliable knowledge base for medical data extraction tasks. Additionally, as a source for lexical and contextual features, Wikipedia was used in the past to improve medical text relation extraction \cite{rink2011automatic}. Guided by prior work,
we build a specialized lexicon by crawling a total of 113 Wikipedia pages under the category of ``Infectious disease'', resulting in 22,123 words representing corpus \textit{M}. Then, we download the same number (22,123) of the most frequent words on all of Wikipedia, representing a general corpus \textit{W}. These can then be used to compute a probability of every word in specialized corpus \textit{M} as: $mp_w=count_w / \sum_w \textit{M}$, as well as the probability of every word in general corpus \textit{W} as $wp_w=count_w / \sum_w \textit{W}$. Next, for every word in every corpus, we compute $p_w=mp_w - wp_w$. Intuitively, the differences in probabilities $p_w$ provide the most descriptive words related to ``infectious disease'' topic which are \emph{not} as prevalent in the general Wikipedia. Ranking the terms by $p_w$, we only keep the top 13,300 meaningful words, as illustrated in Table \ref{Medical_lexicon} (note the topmost words are more specific, while those further down the ranking are more general).
 
\indent
Additionally, Wikipedia references are considered trusted citations as Wikipedia increasingly includes references with high impact factor medical journals such as the \emph{New England Journal of Medicine}, \emph{The Lancet}, the \emph{Journal of the American Medical Association}, and the \emph{British Medical Journal} among the 10 most frequently cited science journals in Wikipedia in 2007 \cite{heilman2011wikipedia}. As Wikipedia pages are usually among the top results of search engine queries \cite{laurent2009seeking, heilman2011wikipedia}, we expect people to use Wikipedia pages and references as a major source of online health information. From the same Wikipedia pages used to collect the medical lexicon, we collect a total of 2,979 referenced URLs from 441 different domains\footnote{Available at http://bit.ly/2m59wpm} including medical literature databases and news agencies. As most Twitter URLs are shortened, we expanded the URLs to detect the original domain.  Finally, we manually classify tweet URL domains as \textit{advocacy} group (advocating specific actions or policies, or claiming to be the best in providing the related information without official ties), \textit{social\_media} (YouTube, Facebook and social media helper websites that forward and aggregate content), \textit{news} (news sources CNN, Reuters, etc.), \textit{informational} (reliable resources providing medical information: medical companies, government sites, Snopes...) or \textit{non-informative} (URLs having no specific domain type). Doing this, we have a total of four different domain type features where every feature is a count of the number of URLs belonging to a domain class in the tweet.

\end{description}

\begin{table}[t]
\centering
\caption{Selected ``Infecious disease'' Wikipedia medical lexicon words}
\label{Medical_lexicon}
\begin{tabular}{|l|cccc|}
\hline
\multicolumn{1}{|c|}{\textbf{Word} (\textbf{\textit{w}})} & \textbf{\textit{mp\textsubscript{w}}} & \textit{\textbf{wp\textsubscript{w}}${}^\bullet$} & \textit{\textbf{p\textsubscript{w}}} & \textbf{Rank} \\ \hline \hline
\textit{syphilis}${}^\ast$                   & 0.01           & -            & 0.01        & 4             \\
\textit{bronchitis}${}^\ast$                 & 0.002          & -            & 0.002       & 81            \\
\textit{tetanus} ${}^\ast$                & 0.001          & -            & 0.001       & 236            \\
\textit{diarrhea}                   & 0.006          & 0.128          & -0.121      & 13682         \\
\textit{epidemiology}               & 0.009          & 0.147          & -0.138      & 15284         \\
\textit{treatment}                  & 0.019          & 4.652          & -4.633      & 33869         \\
\textit{life}                       & 0.003          & 34.61          & -34.608     & 35074         \\ \hline
\multicolumn{5}{p{0.8\linewidth}}{
 ${}^\ast$ Among the chosen top 13,300 words with highest \textit{\textbf{p\textsubscript{w}}} \newline
 ${}^\bullet$ - : is when \textbf{\textit{w}} is not in the \textit{W} corpus
 }
\end{tabular}

\end{table}

In order to pick the best features for the classification task, we employ two different automatic feature selection techniques. First, Information Gain (IG) which is a popular filtering approach that aims at removing irrelevant features after computing the gain value (amount of information a feature brings to the training set) \cite{featureselection}. Second, we use Greedy backward elimination technique (GBE) that starts with a model having all features, and removes features one at a time until reaching a certain performance threshold \cite{featureselection}.

Table \ref{feature-selection} shows the top features each technique produced. Here, we list the top ten features by information gain value and GBE results selecting the best ten features. Based on both techniques, the most significant features correspond to the medical features (advocacy domains count, Wikipedia domains count) followed by the syntax of the tweet text (question marks, exclamation marks...) and the sentiment features (sentiment score, count positive/negative words) and some Twitter features. 

\begin{table}[t]
\centering
\caption{The features selected using Information Gain and Greedy Backward Elimination.}
\label{feature-selection}
{\fontsize{8pt}{9pt}\selectfont
\begin{tabular}{|l|cccc|}
\hline
\textbf{Feature} \Tf & \textbf{min, max} & \textbf{$\mu$ ($\sigma$)} &  \textbf{IG}${}^\star$ & \textbf{GBE}${}^\bullet$\\ \hline \hline
(T) AGE      \Tf   & 61, 281         & 188 (71)             & 9 & \checkmark   \\ 
(T) HAS MENT       & 0, 1  & 0.177 (0.381)  & 10 & \checkmark \\ 
(T) COUNT RT     & 1, 2457         & 394 (713)            & 6 & \checkmark          \\
(S) SENTIMENT      & -2.2, 1.6       & -0.332 (0.71)        & 8 & \checkmark          \\ 
(S) NEG COUNT      & 0, 13           & 0.639 (0.871)        & - & \checkmark          \\ 
(L) HAS QUEST      & 0, 1         & 0.193 (0.395)        & 4 & \checkmark          \\ 
(L) HAS EXCL       & 0, 1            & 0.023 (0.161)        & 5 & - \\ 
(L) VERB CNT       & 0, 38   & 0.673 (0.716) & - & \checkmark \\
(L) ADVB CNT       & 0, 102 & 0.682(0.936)   & 3 & - \\
(L) MULT. '?/!'    & 0, 1            & 0.014 (0.12)         & 2 & \checkmark          \\ 
(M) ADVOCACY CNT   & 0, 2           & 0.045 (0.21)        & 1 & \checkmark          \\ 
(M) WIKI CNT       & 0, 1            & 0.253 (0.435)         & 7 & \checkmark          \\ 

\hline
\multicolumn{5}{p{0.95\linewidth}}{
${}^\star$ \Tf Features are ranked desc according to information gain values.\newline
${}^\bullet$ \checkmark: is in GBE best 10 feature subset, otherwise not.
}
\end{tabular}}
\end{table}

Note that advocacy feature domain type is the strongest feature with high IG value (table \ref{feature-selection}). It is understandable this feature would be useful, given that it requires expert annotation. Further, we find that out of the URLs cited in rumor tweets, 35.0\% were from advocacy websites, 0.1\% from social media, 39.1\% were news and 25.9\% were informative domains, compared to 3.1\% from advocacy and 0.6\% social media, 32.3\% news, and 64.0\% informational in non-rumors, making the presence of advocacy groups and informational sources the distinctive features, and, interestingly, not the news media. Wikipedia domains features is also among the top selected features in both techniques and this features is automatically computed and can be used more broadly.

Finally, we train a supervised classifier to predict which tweets contain rumor and which do not. We build a classifier separately for the top 10 features of IG and GBE techniques. We experiment with three different learning algorithms. First, Na{\"i}ve-Bayes algorithm \cite{zhang2004naive}, a probabilistic based on Bayes' theorem with strong (``na{\"i}ve'') independence assumptions between the features. Second, Random Forest \cite{ho1995random}, which is a collection of classifiers where every classifier votes for one class and every instance is classified based on the majority class. Third, Random Decision Tree \cite{du2002building}, a classifier that recursively builds a tree by splitting the training data based on a criterion until all partitions have the same class label. 
We find the best results using the Random Tree classifier using the top 10 GBE features. For training/validation process, we perform 10-fold cross validation, in which 10 experiments are performed on a different tenth of the data held out for testing, such that we take advantage of the whole dataset for both training and testing. A summary of the best classifier (Random Tree) with top 10 GBE features results are shown in Table \ref{classifier-result}. As it shows, the classifier achieves a precision of 0.946 with recall 0.944 which is significantly better than a random predictor. The F-value (a harmonic mean of precision and recall) is high, indicating a good balance between precision and recall values. The final row presents the average values from across both classes. Note that these results are overfitted, given the limited amount of data available, feature selection on the test set, and also that the method relies on manually labeled tweets, with the addition that the dataset is already topically specialized. 


\begin{table}[t]
\centering
\caption{Classification performance on the Rumor vs. non-Rumor task, using Random Tree classifier with GBE features. }
\label{classifier-result}
{\fontsize{8pt}{9pt}\selectfont
\begin{tabular}{|l|c|c|c|}
\hline

\textbf{Class}\Tf   & \textbf{Precision} & \textbf{Recall} & \textbf{F-measure} \\ \hline \hline
rumor      & 0.929 & 0.921 \Tf  & 0.925   \\
non-rumor  & 0.963 & 0.967   & 0.965 \\
weighted average    & 0.946  & 0.944   &   0.945 \\\hline
\end{tabular}
}
\end{table}

As we find having training data within the topic to be extremely helpful in building accurate classifiers, we explore a more challenging scenario wherein the classifier is trained on 5 topics and tested on the 6th. The results are shown in Table \ref{Topic classification}. Every row of the table shows which topic is excluded in training the classifier, and then is used for testing. We find the performance is not uniform, with topics 1 and 5 having the worst precision, while topics 3 and 4 having recall under 0.500. Once again, this points to the importance of expert labeled data that is topically matched to the one in question.

\begin{table}[h]
\centering
\caption{Classification performance of detecting rumor tweets in individual topics, using Random Tree classifier with GBE features.}
\label{Topic classification}
\resizebox{\columnwidth}{!}{%
\begin{tabular}{|l|c|c|c|c|}
\hline
\textbf{Topic} & \textbf{\% Rumor} & \textbf{Precision} & \textbf{Recall} & \textbf{F-Measure}  \\ \hline \hline
R1. Zika linked to GMO                  \Tf   & 61\%     & 0.296   & 0.869   & 0.440    \\ \hline
R2. Flu symptoms similar to Zika        \Tf   & 31\%     & 0.746   & 0.504   & 0.602    \\ \hline
R3. Vaccines cause microcephaly         \Tf   & 71\%     & 0.683   & 0.490   & 0.571    \\ \hline
R4. Insecticide cause microcephaly      \Tf   & 32\%     & 0.594   & 0.432   & 0.500    \\ \hline
R5. Americans are immune to Zika        \Tf   & 32\%     & 0.101   & 0.523   & 0.170    \\ \hline
R6. Coffee as mosquito repellent        \Tf   & 31\%     & 0.688   & 0.688   & 0.688    \\ \hline

\end{tabular}%
}
\end{table}

%% file: discussion.tex

\section{Discussion}
\label{sec:discussion}

\textbf{Key Findings.}

Communication on Twitter around a major public health crisis is an essential component in the public health response. In our search of rumors in the stream of Zika-related tweets, we find automatic topic discovery tools such as LDA to be too coarse-grained to tease out the rumors WHO and Snopes have cited as most concerning. Thus, we incorporate the expert knowledge to compose high-precision queries to retrieve the relevant tweets. We also show that further steps are needed, as after a closer examination we find only roughly half of the captured tweets to be actual rumors. This insight shows the perils of using keyword or hashtag-based topic definition, as is done, for example by ``Truthy'' \cite{ratkiewicz2011truthy} where a topic is defined by a single hashtag, or even in Castillo et al.\  \cite{castillo2011information} who use Twitter Monitor algorithm to formulate keyword-based queries.

Further, within the small sample of topics we examined, we discovered a variety in terms of longevity. Topics relating to everyday activities, such as seasonal flu or coffee, can be a subject to hyperbole and humor which may propagate the misinformation. However, rumors originated from known advocacy websites such as http://www.march-against-monsanto.com/ may display a spike which quickly dissipates without correction. These websites adjust their stances to the new trending topics like Zika while maintaining their core message. 

Interestingly, mainstream news websites were cited at roughly the same rate in rumor tweets (39.1\%) as in others (32.3\%), including the clarifications. This emphasizes the importance of authoritative sources outside mainstream news media in setting the record straight. Further, Towers et al.\ \cite{towers2015mass} find that mainstream news media may help spread fear and misinformation, such as in the case of Ebola in 2014, ``with each Ebola-related news video inspiring tens of thousands of Ebola-related tweets and Internet searches'', effectively spreading unsubstantiated panic in the United States.



\textbf{Public Health Relevance.}

Detecting health rumors in a timely fashion can help public health officials tackle them before they spread. However, over-reacting to a rumor might in fact increase its damage by advertising the harmful misconceptions. In the case of the Ebola outbreak, some of the rumors circulated on the Internet, such as that drinking salty water was an effective protective measure, led to several deaths \cite{Oyeyemi2014,Jin2014}. Rumors around a vaccination trial for a new Ebola vaccine sparked fears for a regular \textit{Measles} vaccine, which was being used to tackle a \textit{Measles} outbreak at the same time \cite{TheVaccineConfidenceProject2015}. Public health decisions in one country can spark rumors and mistrust in another, such as when the HPV (\textit{Human Papilloma} Virus) vaccine campaign discontinuation in Japan sparked concerns and rumors about its safety worldwide \cite{Larson2014}. Thus, it is imperative that the impact of health-awareness campaigns is monitored in real time, as well as internationally. The tools described here can help public health practitioners in tackling the large scale of social media streams.

Further, the case of Zika is highly complex, as much uncertainty surrounded important information. For instance, the pathogenesis of microcephaly took months to be established. Previous works have highlighted the difficulty of early detection of rumors (i.e. \cite{Zhao2015}) in public health cases -- to assess the veracity of a rumor can take months of public health investigations. However, due to the unprecedented scale of the crisis, health authorities started to act before a clear link between the Zika outbreak and microcephaly was established. This was especially challenging, since it can happen that apparent rumors are in fact truth. As an example, reports on narcolepsy as a side effect of a flu vaccine in the Nordic countries were first depicted as rumors, but later, few cases were confirmed and that took years of research and still it is contested \cite{Sturkenboom2015}. Although correlation was found in epidemiological data, some scholars argue that an increase in awareness due the hype of the ``vaccination crisis'' might have caused the increase in cases. Public health authorities are continuously working in a complex crisis communication dilemma, since they have to act on some level of uncertainty. In this study, we chose the rumors which have been identified by authoritative sources as certain. However, a different approach may be called for the detection of possible health rumors, which is an exciting future research direction. 


In this context, we believe more work is needed in the integration of rumor monitoring with public health officials, and especially the work-flow of communication departments of public health authorities. A pipeline such as AIDR (which provided our collection) described by Imran et al.\ \cite{imran2014aidr} wherein volunteers provide labeled social media during a disaster to train automated methods, may also be useful for ongoing health emergencies.

\textbf{Limitations.} 

One of the main challenges of this study is that we cannot be sure about the representativeness of the social media users compared with the general population. The demographics of social media users tend to be young, and female \cite{duggan2013demographics}, which may be important, as some have called women ``gatekeepers'' of their families health \cite{calabretta2002consumer,warner2004toward}. In addition, we need to consider that particular segments of the population are more at-risk (e.g. pregnant women) and it may be difficult to identify such users online (however, tracking this particular group of users would enlighten the effect Zika has on child-baring women). Further, the limited resources of this study were applied to only a handful of rumors -- those especially brought up by WHO and Snopes -- and a closer collaboration with health communication experts may provide further insight into the variety of misinformation both online and its interaction with mainstream media. Finally, Zika affected many countries, and our original dataset has covered several languages. The peculiarities of rumors in each language (and by proxy, in perhaps different cultures), could illuminate differences in the perception of medical information on social media.


%% file: conclusion.tex

\section{Conclusion}
\label{sec:conclusion}

This paper presents tool pipeline incorporating expert knowledge, crowdsourcing, and machine learning for health-related rumor discovery in a social media stream. Each step of the analysis was rigorously tested by manual evaluation, providing qualitative and quantitative insight into a process needed to collect data relevant to the health communication professionals. 

In particular, our study shows that tracking health misinformation in social media is not trivial, and requires some expert supervision. This can then augmented by ``crowd'' workers in order to provide additional annotation of the captured rumor-related tweets. We show the bursty and varied nature of the Zika rumors, some provoked by known advocacy groups, others propagated due to their affordance for humor or light banter. We find traditional media sources not to be prominent in clarifying rumors, but instead show the importance of authoritative informational sources. We hope this work will encourage a collaboration between health professionals and data researchers in order to quickly understand and mitigate health misinformation on social media.

Continued work will address the multi-lingual nature of the dataset, and expand the efforts to cross-language analysis of rumors and their potential international spread. More studies on health rumors may provide richer test beds for building automatic classifiers not just for rumors, but for the detection of informational campaigns. Finally, a user-friendly interface similar to Kostkova et al.\  \cite{kostkova2016vac}, which may involve expert input, like AIDR \cite{imran2014aidr}, would smooth the interaction between data scientists and health communication professionals.